\documentclass[aps,prc,preprint,showpacs,showkeys,floatfix,nofootinbib,superscriptaddress,a4paper]{revtex4}

\usepackage{epsfig,amssymb,amsmath}
\usepackage{bm}
\newcommand{\Grad}{$^o$}

\begin{document}
\thispagestyle{empty}
\title{Isotopic Production Cross Sections in Proton-Nucleus
Collisions at 200 MeV}
\author{H. Machner}
\email[]{h.machner@fz-juelich.de} \affiliation{Institut f\"{u}r
Kernphysik, Forschungszentrum J\"{u}lich, 52425 J\"{u}lich, Germany
and Dept. of Physics, University of the Witwatersrand, Johannesburg,
South Africa}

\author{D. G. Aschman}
\affiliation{Dept. of Physics, University of Cape Town,
Johannesburg, South Africa}

\author{K. Baruth-Ram}
\affiliation{Dept. of Physics, University of Durban-Westville, South
Africa}

\author{J. Carter}
\affiliation{Dept. of Physics, University of the Witwatersrand,
Johannesburg, South Africa}

\author{A. A. Cowley}
\affiliation{Dept. of Physics, University of Stellenbosch,
Stellenbosch, South Africa}

\author{F. Goldenbaum}
\affiliation{Institut f\"{u}r Kernphysik, Forschungszentrum
J\"{u}lich, 52425 J\"{u}lich, Germany}

\author{B. M. Nangu}
\affiliation{Dept. of Physics, University of Zululand,
Kwa-Dlangezwa, South Africa}

\author{J. V. Pilcher}
\affiliation{iThemba Labs, Faure, South Africa}

\author {E. Sideras-Haddad}
\affiliation{Dept. of Physics, University of the Witwatersrand,
Johannesburg, South Africa}

\author{J. P. F. Sellschop}
\thanks{deceased}
\affiliation{Dept. of Physics, University of the Witwatersrand,
Johannesburg, South Africa}

\author{F. D. Smit}
\affiliation{iThemba Labs, Faure, South Africa}

\author{B. Spoelstra}
\affiliation{Dept. of Physics, University of Zululand,
Kwa-Dlangezwa, South Africa}

\author{D. Steyn}
\affiliation{Dept. of Physics, University of Cape Town,
Johannesburg, South Africa}

\date{\today}

\begin{abstract}
Intermediate mass fragments (IMF) from the interaction of $^{27}$Al,
$^{59}$Co and $^{197}$Au with 200 MeV protons were measured in an
angular range from 20 degree to 120 degree in the laboratory system.
The fragments, ranging from isotopes of helium up to isotopes of
carbon, were isotopically resolved. Double differential cross
sections, energy differential cross sections and total cross
sections were extracted. 
\end{abstract}
\maketitle
\section{Introduction}\label{sec:Introduction}
Studies of spallation processes, both experimental and theoretical,
are numerous. One reason for this may be the importance of knowledge
of cross sections and reaction mechanisms for our understanding of
cosmic rays \cite{Auger39, Brown49, Powell59, Fields00, Biermann01,
Hoerandel01, Nolfo01} and the production of cosmogenic radionuclides
\cite{Michel96, Waddington99}, and the process of neutron production
in spallation sources. Recent reviews of the process can be found in
Refs \cite{Huefner85, Silberberg90}. Most of the experimental data
exist in the range above 1 GeV which is important for spallation
neutron source construction and the understanding of very high
energy cosmic rays. However, the energy of the maximum abundance of
protons in cosmic rays is around 200 MeV \cite{Austin81, Simpson83}.
We measured intermediate mass fragments at a proton beam energy of
200 MeV incident on three targets spanning the periodic table,
namely, $^{27}$Al, $^{59}$Co and $^{197}$Au. These data complement
previous cross sections for proton, deuteron and tritium emission on
$^{27}$Al and $^{197}$Au \cite{Didelez82, Machner84}. The cross
sections given there for $^3$He and $\alpha$-particles are too small
when compared to systematics \cite{Kalbach87,Kalbach88}. They were
measured with a set up different to that used for the hydrogen
isotopes and might be low by a factor of 4. We will come back to
this point. They also complement data for a silver target taken at
proton energies close by \cite{Green80, Yennello90}.

\section{Experiments}\label{sec:Experiments}
The experiment was performed at the separated-sector cyclotron
facility of iThemba labs. A detailed description of the layout of
the facility and equipment is given in Ref. \cite{Pilcher89} and
references therein.  The beam of 200 MeV was focused to a spot size
of less than 2 $\times$ 2 mm at the target center of a 1.5 m
diameter scattering chamber. Great care was taken to minimize the
halo of the incident proton beam by focusing the beam through a 3 mm
diameter hole in a ruby scintillator target. The targets were self
supporting foils with thicknesses of 2.9~mg/cm$^2$, 1.0~mg/cm$^2$
and 4.0~mg/cm$^2$ for $^{27}$Al, $^{59}$Co and $^{197}$Au,
respectively. The target materials had purity of 99.9$\%$. A
possible (invisible) oxidation of the surface in case of the
aluminum target leads to a negligible amount of oxygen. Fragments
were measured with a telescope consisting of an active collimator
followed by three silicon detectors with thicknesses of 50~$\mu$m,
150~$\mu$m and 1 mm. The solid angle of the telescope was 2.2 msr.
Another 1 mm thick detector vetoed penetrating hydrogen and helium
isotopes. The detectors were calibrated with radioactive sources and
a precision pulse generator. In order to reduce electronic noise
they were cooled to a few degrees with chilled water. Detection
angles were from 20\Grad to 120\Grad. The opening angle of the
collimator resulted in an angle uncertainty of $\pm 2.2$\Grad. The
incident proton flux was measured by a beam dump Faraday cup.

The $\Delta E-E$ method was used for particle identification. A
linearized particle identification quantity, $PI$, was obtained from
the energy-range relation, given by
\begin{equation}\label{equ:de2}
PI=\left[\left(E_1+E_2\right)^b-E_2^b\right]/d_1
\end{equation}
if the particle is stopped in the second detector.  $E_i$ denotes
the energy deposited in the i-th detector and $d$ is the detector
thickness. If the particle is stopped in the third detector one has
the relation
\begin{equation}\label{equ:de3}
PI=\left[\left(E_1+E_2+E_3\right)^b-E_3^b\right]/(d_1+d_2).
\end{equation}
Furthermore
\begin{equation}
mZ^2\propto PI.
\end{equation}
with $m$ the fragment mass and $Z$ the charge number.
\begin{figure}[h]
\includegraphics[width=8 cm]{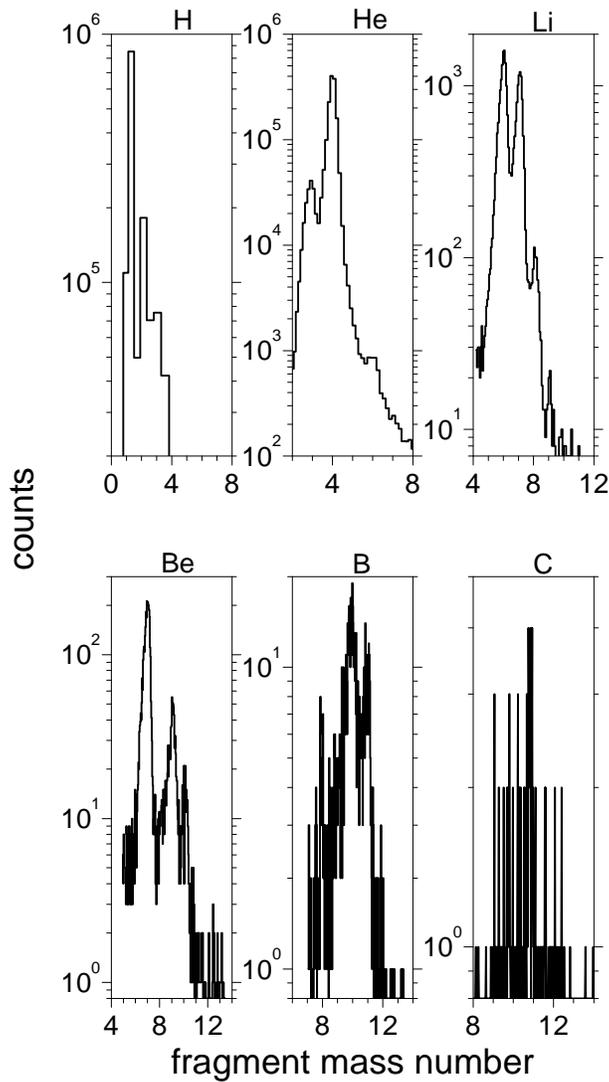}
\caption{Mass distributions from the interaction of 200 MeV protons
with $^{59}Co$ measured at 20\Grad.} \label{fig:PIO}
\end{figure}
For the exponent a value $b=1.73$ was used. As an example a mass
distribution, obtained by dividing the ranges in the PI-spectrum by
$Z^2$, is shown in Fig. \ref{fig:PIO} for the case of cobalt.
Hydrogen isotopes fulfilling the energy conditions (\ref{equ:de2})
or (\ref{equ:de3}) have the largest yield, but are not considered
here. Good isotope separation is visible up to boron. In the case of
the gold target even carbon fragments could be resolved.

The counting rate was then converted to cross sections. The
following systematic errors contribute to the total uncertainty. The
target thicknesses are known with typically 10$\%$ uncertainty. The
incident flux was measured with 2$\%$ uncertainty, while the solid
angle, electronic dead time correction and energy calibration were
estimated to contribute in total to less than 2$\%$. The emission
angles are uncertain to $\pm 2.2$\Grad. The error bars in the
figures show only the statistical uncertainty.

\begin{figure}[h]
\begin{center}
\includegraphics[width=7 cm]{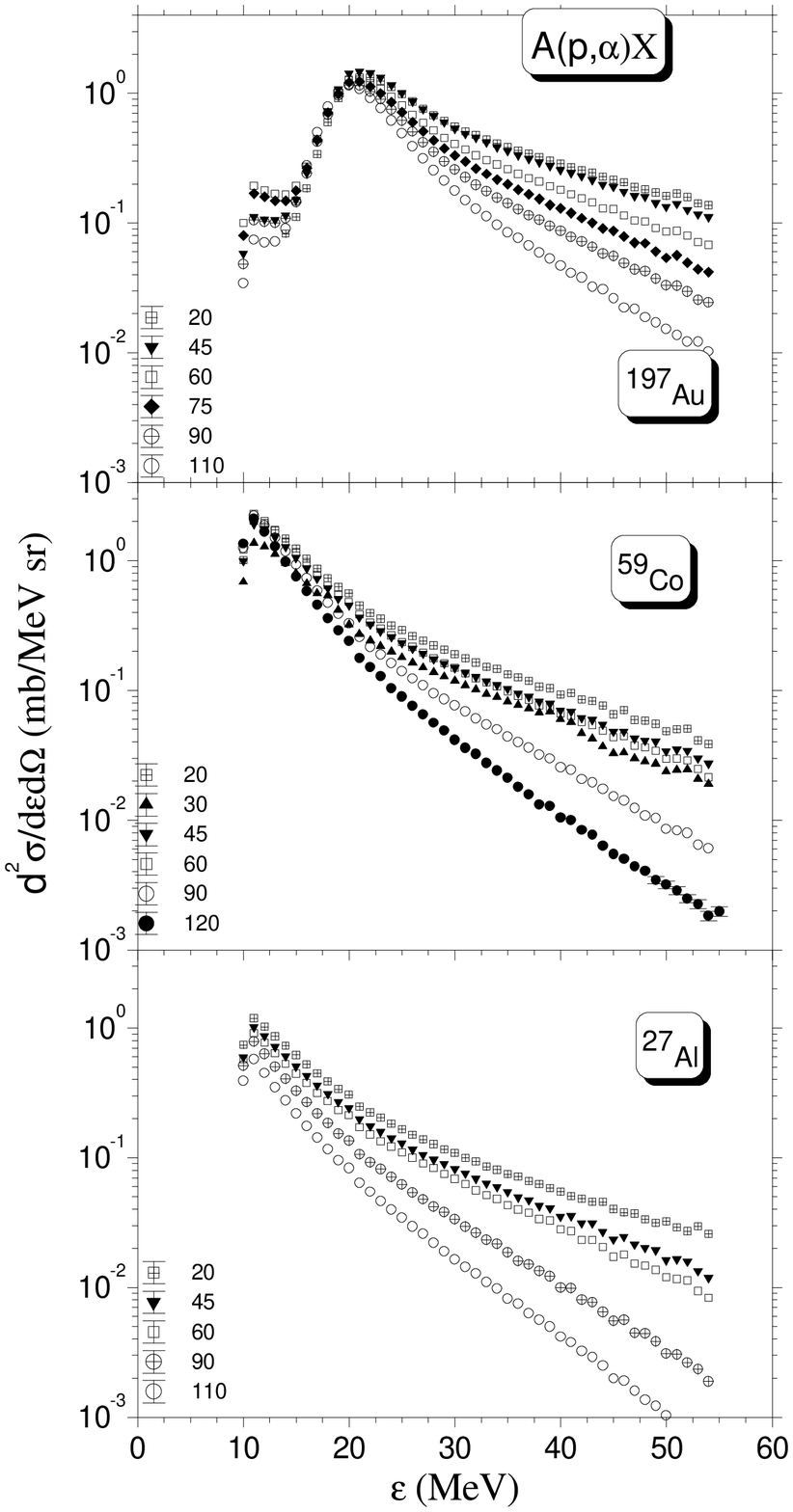}
\caption{Energy spectra of $\alpha$ particles for the given angles
and the given targets.} \label{He4_dd}
\end{center}
\end{figure}

\begin{figure}[h]
\begin{center}
\includegraphics[width=7 cm]{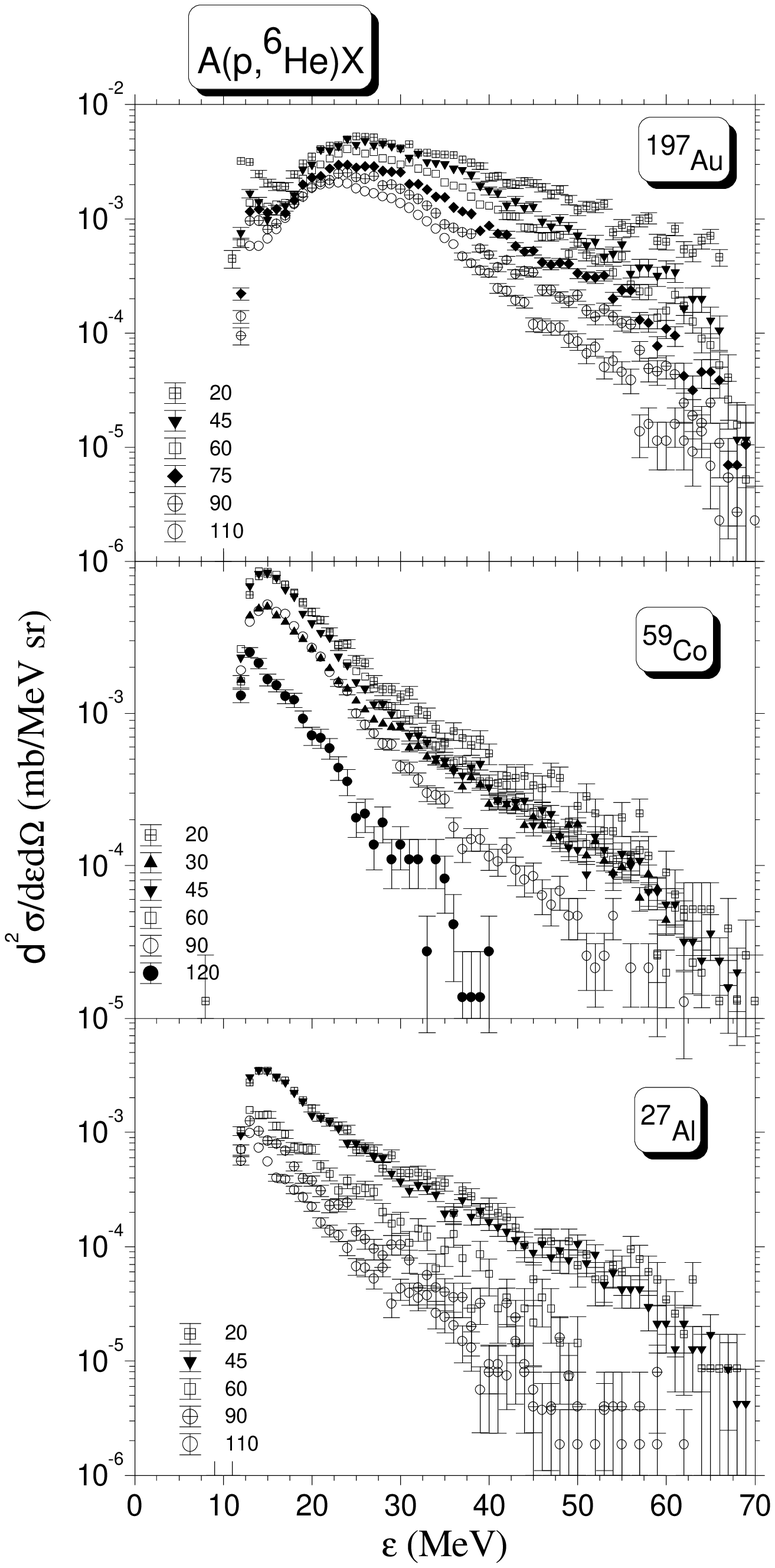}
\caption{Same as Fig. \ref{He4_dd} but for $^6$He emission.}
\label{He6_dd}
\end{center}
\end{figure}

\begin{figure}[h]
\begin{center}
\includegraphics[width=7 cm]{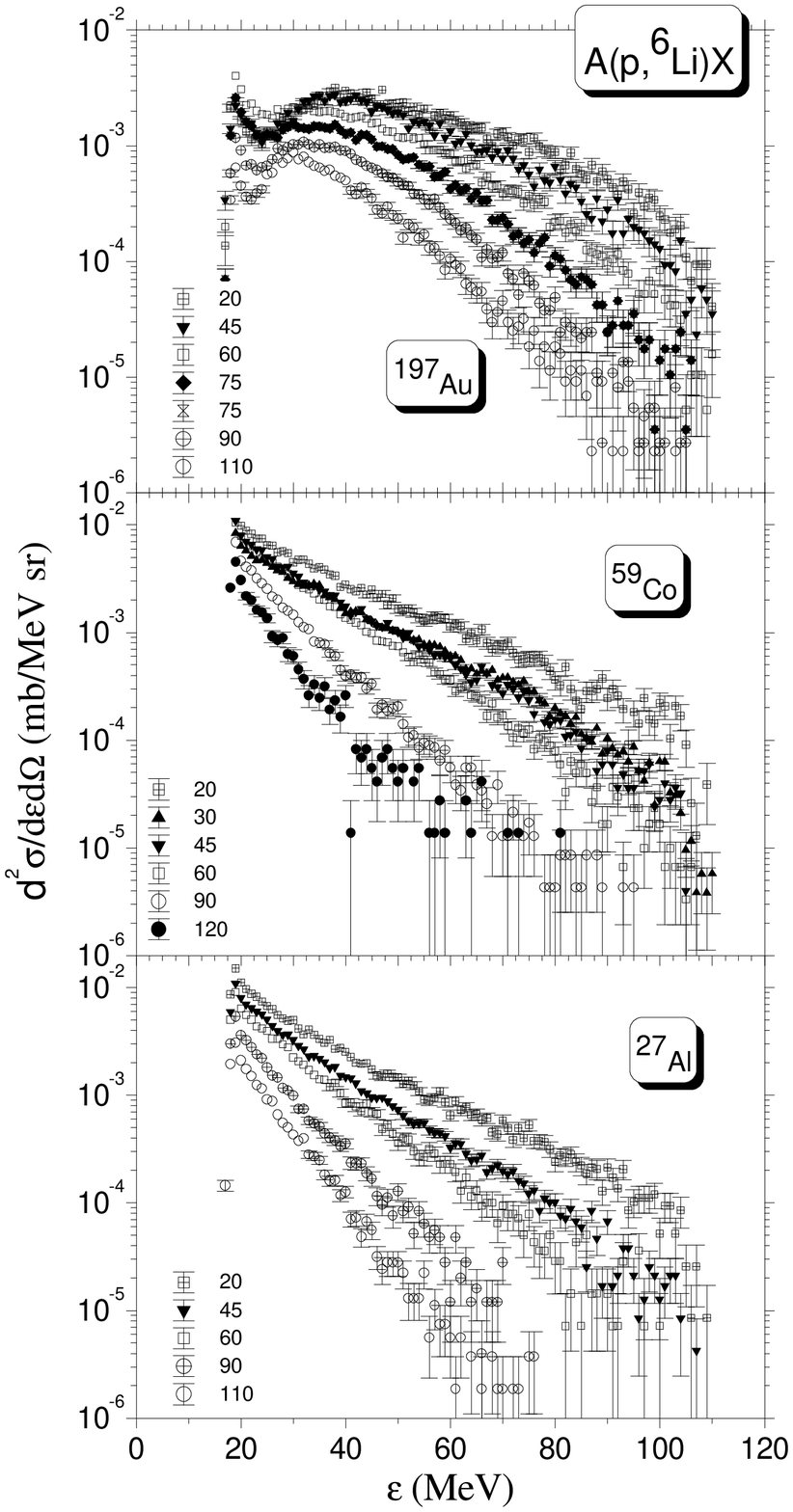}
\caption{Same as Fig. \ref{He4_dd} but for $^6$Li emission.}
\label{Li6_dd}
\end{center}
\end{figure}

\begin{figure}[h]
\begin{center}
\includegraphics[width=7 cm]{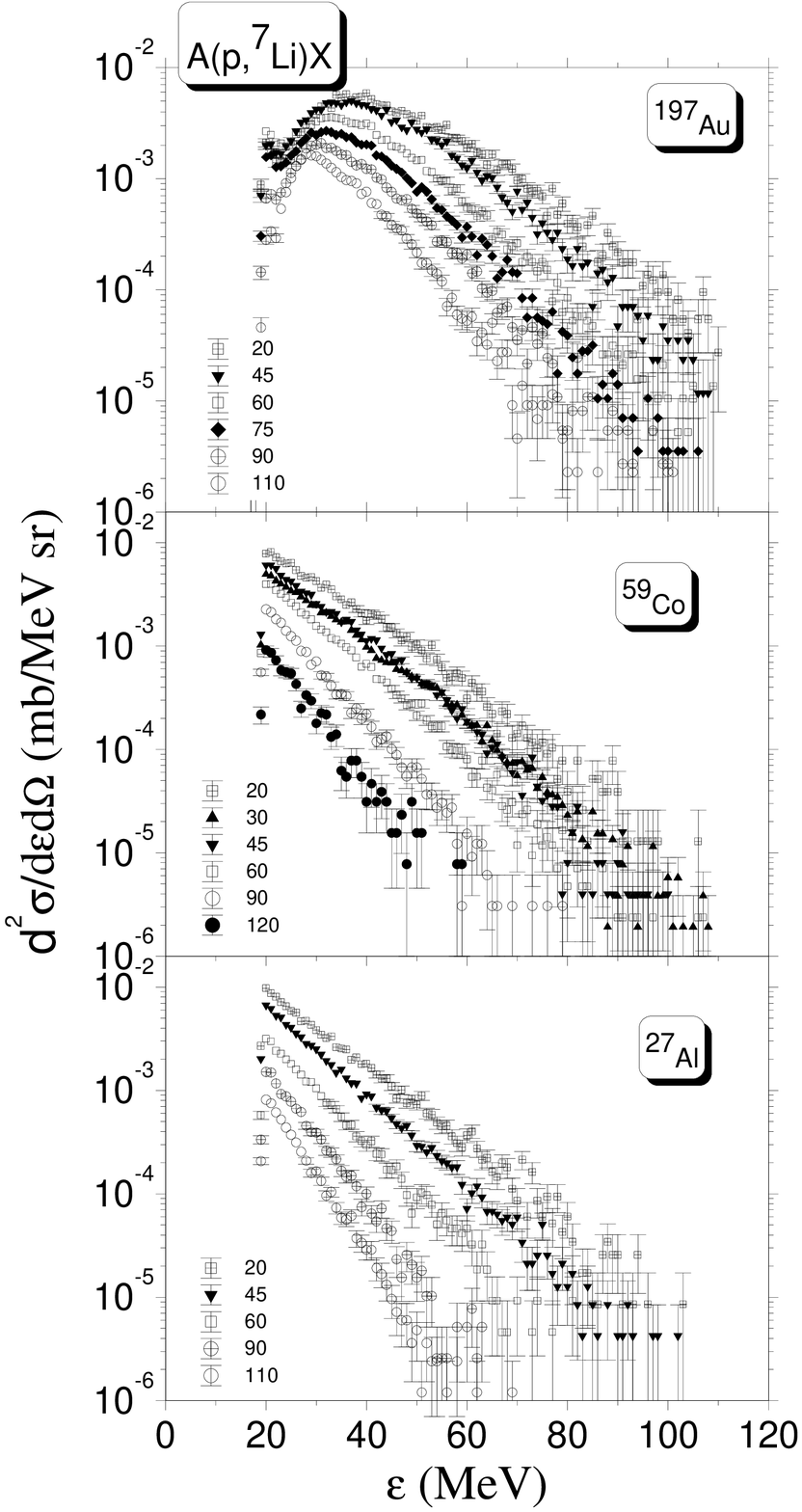}
\caption{Same as Fig. \ref{He4_dd} but for $^7$Li emission.}
\label{Li7_dd}
\end{center}
\end{figure}

\begin{figure}[h]
\begin{center}
\includegraphics[width=7 cm]{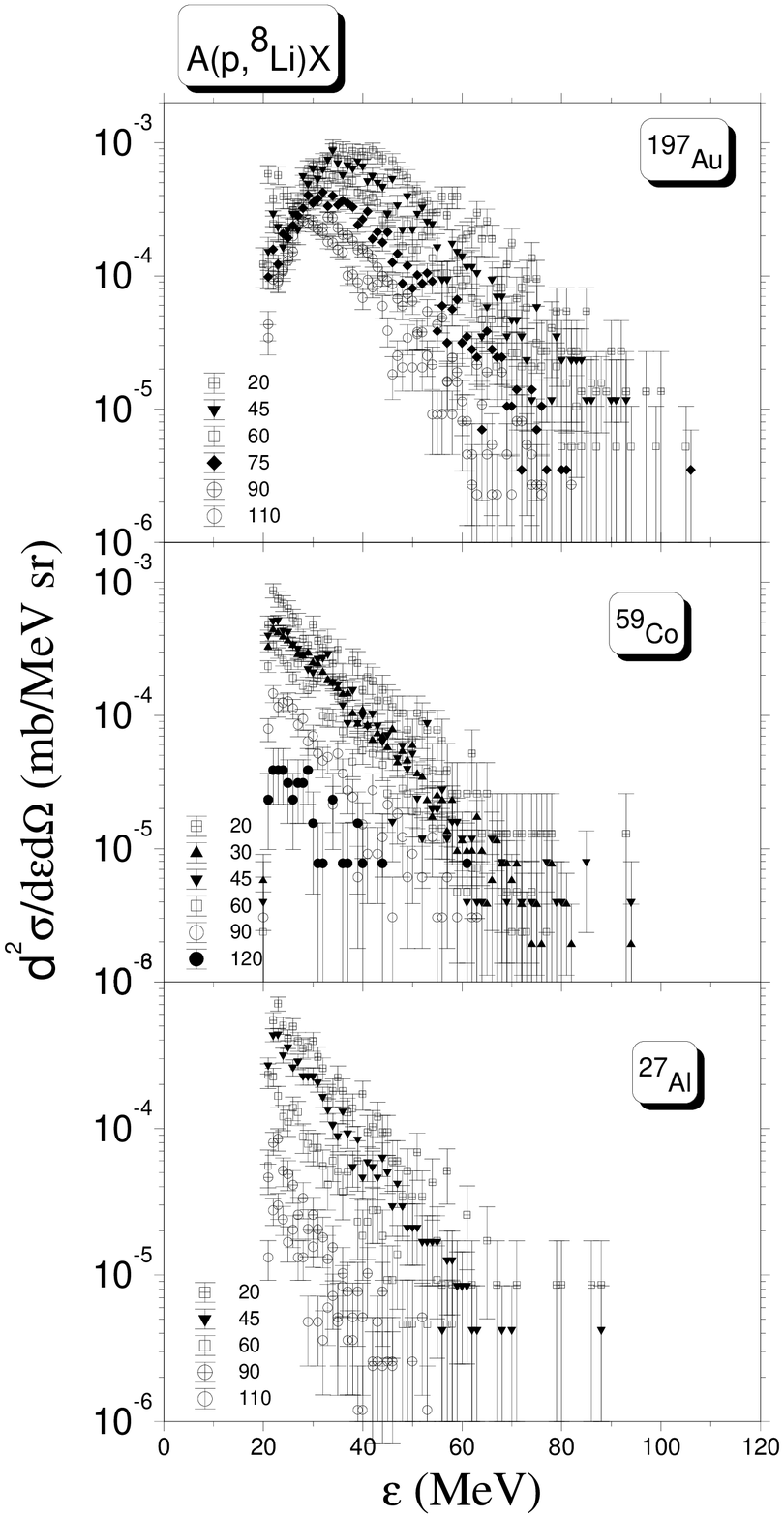}
\caption{Same as Fig. \ref{He4_dd} but for $^8$Li emission.}
\label{Li8_dd}
\end{center}
\end{figure}

\begin{figure}[h]
\begin{center}
\includegraphics[width=7 cm]{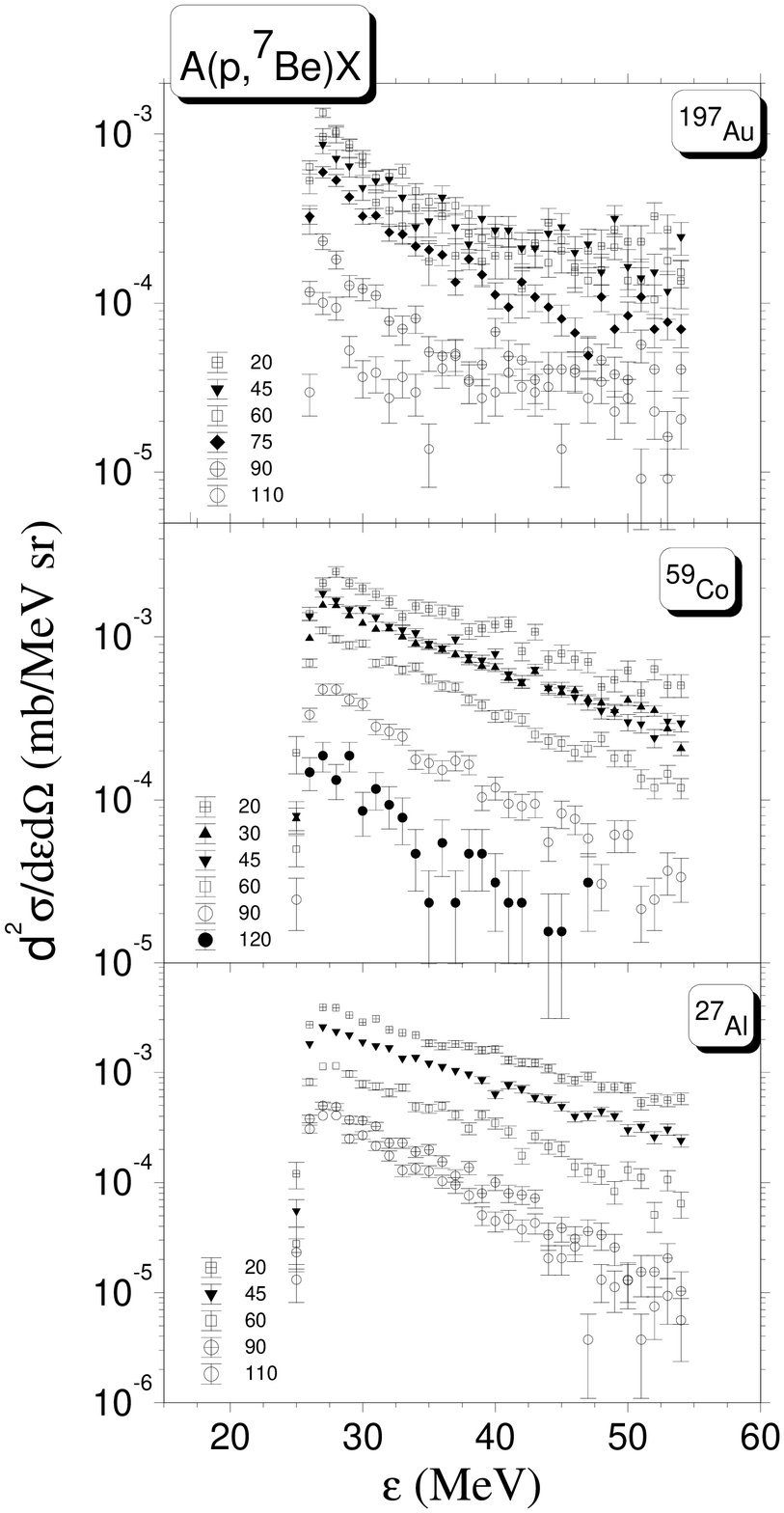}
\caption{Same as Fig. \ref{He4_dd} but for $^7$Be emission.}
\label{Be7_dd}
\end{center}
\end{figure}

\begin{figure}[h]
\begin{center}
\includegraphics[width=7 cm]{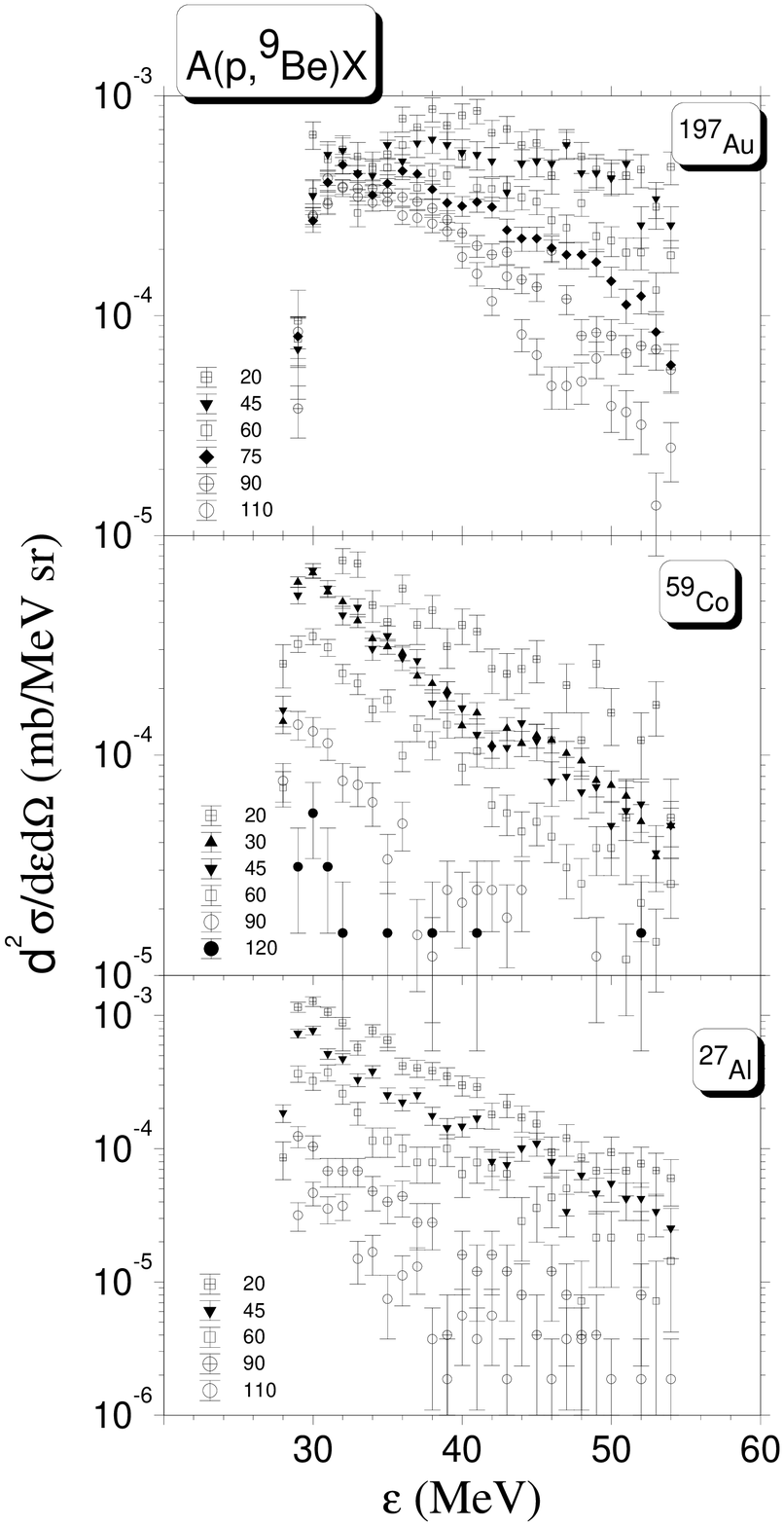}
\caption{Same as Fig. \ref{He4_dd} but for $^9$Be emission.}
\label{Be9_dd}
\end{center}
\end{figure}

\begin{figure}[h]
\begin{center}
\includegraphics[width=7 cm]{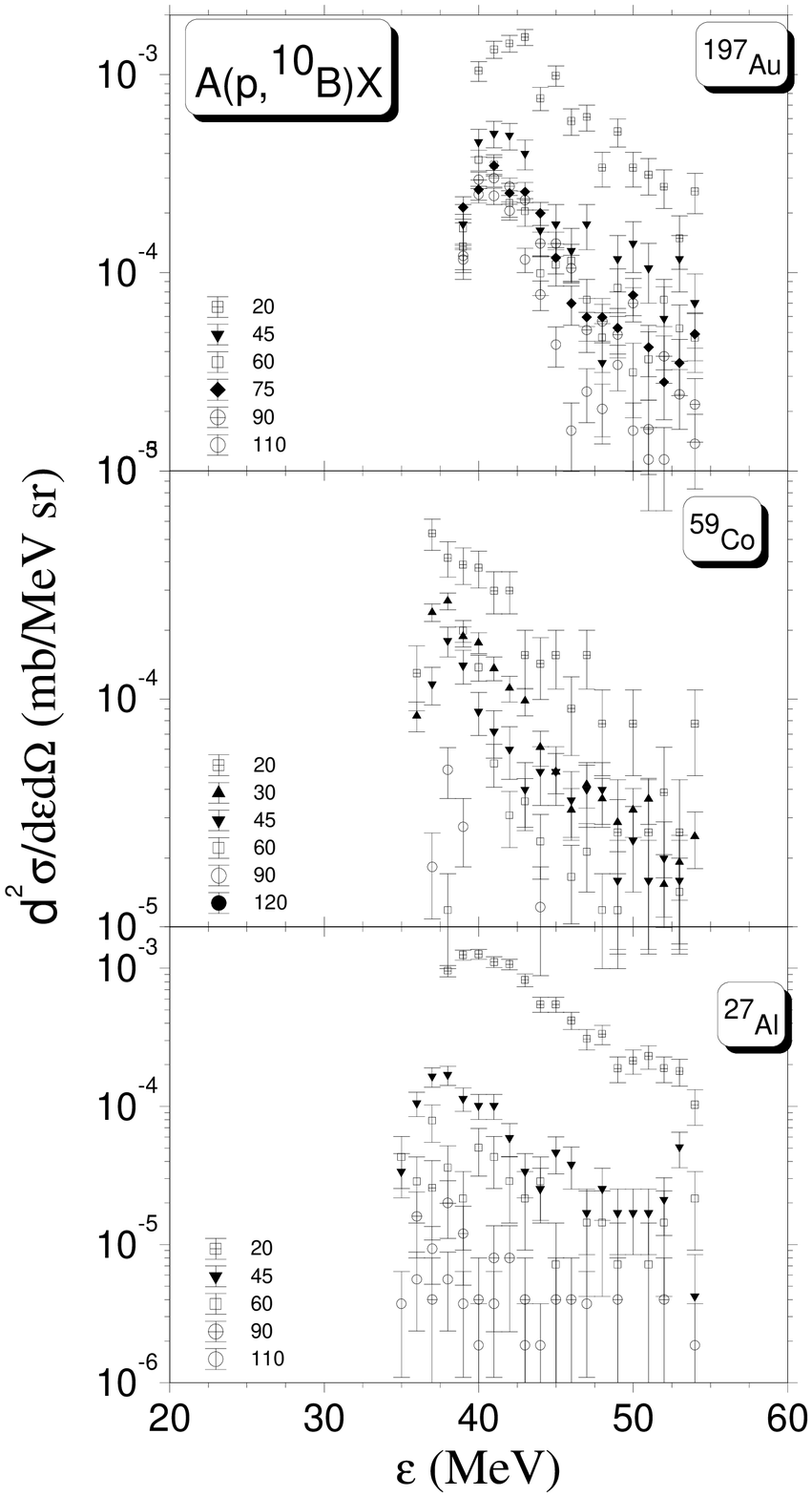}
\caption{Same as Fig. \ref{He4_dd} but for $^{10}$B emission.}
\label{B10_dd}
\end{center}
\end{figure}

In Figs \ref{He4_dd}-- \ref{B10_dd} double differential cross
sections are shown for IMFs ranging from $^4He$ to $^{10}B$. The
statistics get poorer with increasing mass number.

The IMF spectra in the case of the gold target show the effect of
the Coulomb barrier: a maximum which is in most cases close to 10
MeV$\times Z$ with $Z$ the fragment charge number. In the case of
the cobalt target this is just at the detection threshold which is
given by the thickness of the first $\Delta E$ detector. For
aluminum the Coulomb barrier is below our detection range. In case
of the gold target a second component shows up. This is emission
below the Coulomb barrier of a gold-like system. It is obviously
emission from a system with a much smaller Coulomb barrier.
Unfortunately, the first $\Delta E$ detector is too thick to study
such a component in the case of the other targets. Such a component
can be explained as emission from fission fragments which in the
case of lighter target nuclei are not as frequent as in the case of
gold and was also seen in the emission of low energy protons
following $\bar p$ absorption on uranium nuclei \cite{Markiel88}.

\section{Data Analysis}\label{sec:Data-Analysis}
Cross sections were analyzed in terms of a simple model assuming a
moving source prescription. For completeness the content of the
model \cite{Machner90} is briefly repeated here. Suppose an IMF is
emitted statistically from a source. The intensity distribution, by
assumption a Maxwell-Boltzmann distribution, is isotropic in the
rest system of the source. In the laboratory system we then have
\begin{eqnarray}\label{eqn:moving_source}
\frac{d^2\sigma(\theta,\epsilon)}{d\Omega d\epsilon} = \nonumber \\
 C\sqrt{\epsilon}
\exp\left[-\left(\epsilon-\sqrt{2m\epsilon} v\cos\theta
+\frac{1}{2}mv^2\right)/T\right]
\end{eqnarray}
where $C$ is a normalization constant, $\theta$ the emission angle
and $\epsilon$ the energy of the fragment.  $m$ is the mass of the
fragment, $v$ denotes the velocity of the source and $T$ its
temperature. In the present model $v=v(\epsilon)$ and
$T=T(\epsilon)$ and not constants as in the usual moving source
model. It is a common belief that in the early stage of a reaction
the excitation energy is shared by a small number of nucleons. Thus
momentum and energy conservation require a large source velocity and
a high temperature in this stage, which is represented by the high
energy of the IMF. At a later stage a succession of nucleon-nucleon
interactions have taken place and more nucleons are in the source.
This results in a smaller source velocity but higher temperature.
How can one extract these two quantities? Unfortunately it is
impossible. One can extract only a function of both quantities. The
logarithm of the cross section is
\begin{eqnarray}\label{eqn:slope}
\ln\left[\frac{d^2\sigma(\theta,\epsilon)}{d\Omega d\epsilon}\right]
=  \ln(C\sqrt{\epsilon})+\nonumber \\
-\left(\epsilon+\frac{1}{2}mv^2-\sqrt{2m\epsilon} v\cos\theta \right)/T \nonumber \\
= a(\epsilon)\cos\theta + b(\epsilon).
\end{eqnarray}
In the last line we have used the abbreviations
\begin{equation}
b(\epsilon)=\ln(C\sqrt{\epsilon})-\left( \epsilon + \frac{1}{2} mv^2
\right)/T
\end{equation}
and
\begin{equation}\label{equ:a_param}
a(\epsilon)=\frac{\sqrt{2m\epsilon}v}{T}.
\end{equation}
We have chosen $a$ and $b$ in such a way to be consistent with
earlier nomenclature \cite{Machner90}. Linear fits to the logarithm
of the double differential cross section versus the cosine of the
emission angle are excellent.
\begin{figure}[h]
\begin{center}
\includegraphics[width=5 cm]{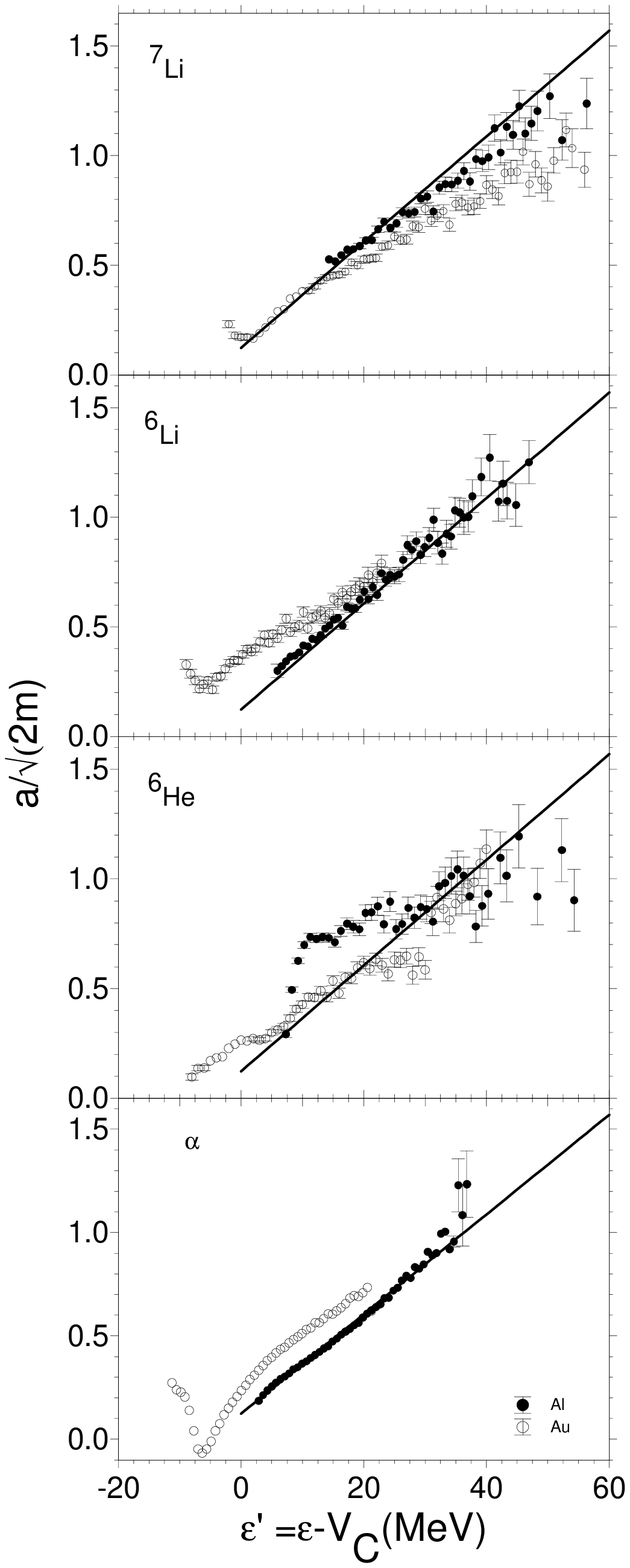}
\caption{The slope parameter as function of the Coulomb corrected
energy.} \label{red_slope_VC}
\end{center}
\end{figure}
Both fit parameters $a$ and $b$ contain the source velocity and the
temperature. Since $b$ also contains the normalization constant it
is impossible to deduce the numbers of interest.

An emitted IMF is accelerated in the Coulomb field. In order to
compare the energies before acceleration we study $a/\sqrt{2m}$ as
function of $\epsilon-V_C$. This is done in Fig. \ref{red_slope_VC}
for the two targets aluminum and gold and for IMFs' $\alpha$'s,
$^6He$'s, $^6Li$'s and $^7Li$'s. These are cases with reasonable or
even good statistics. In the case of gold two components are
visible: that for the higher energies is a smooth curve while for
smaller energies $a$ reflects the barrier penetration. It is
interesting to note that the higher energy component follows almost
a straight line with uniform slope. In order to show this effects we
have fitted a straight line to the case of $\alpha$-particle
emission with the aluminum target. This line without any shift also
passes through the bulk of points in case of other IMF types
although not perfectly fitting the data. This may be an evidence
that equilibration proceeds in an almost unique way independent of
the target size. However, in order to extract angle integrated cross
sections the fitted values with error bars were applied.

\begin{figure}[h]
\includegraphics[width=8cm]{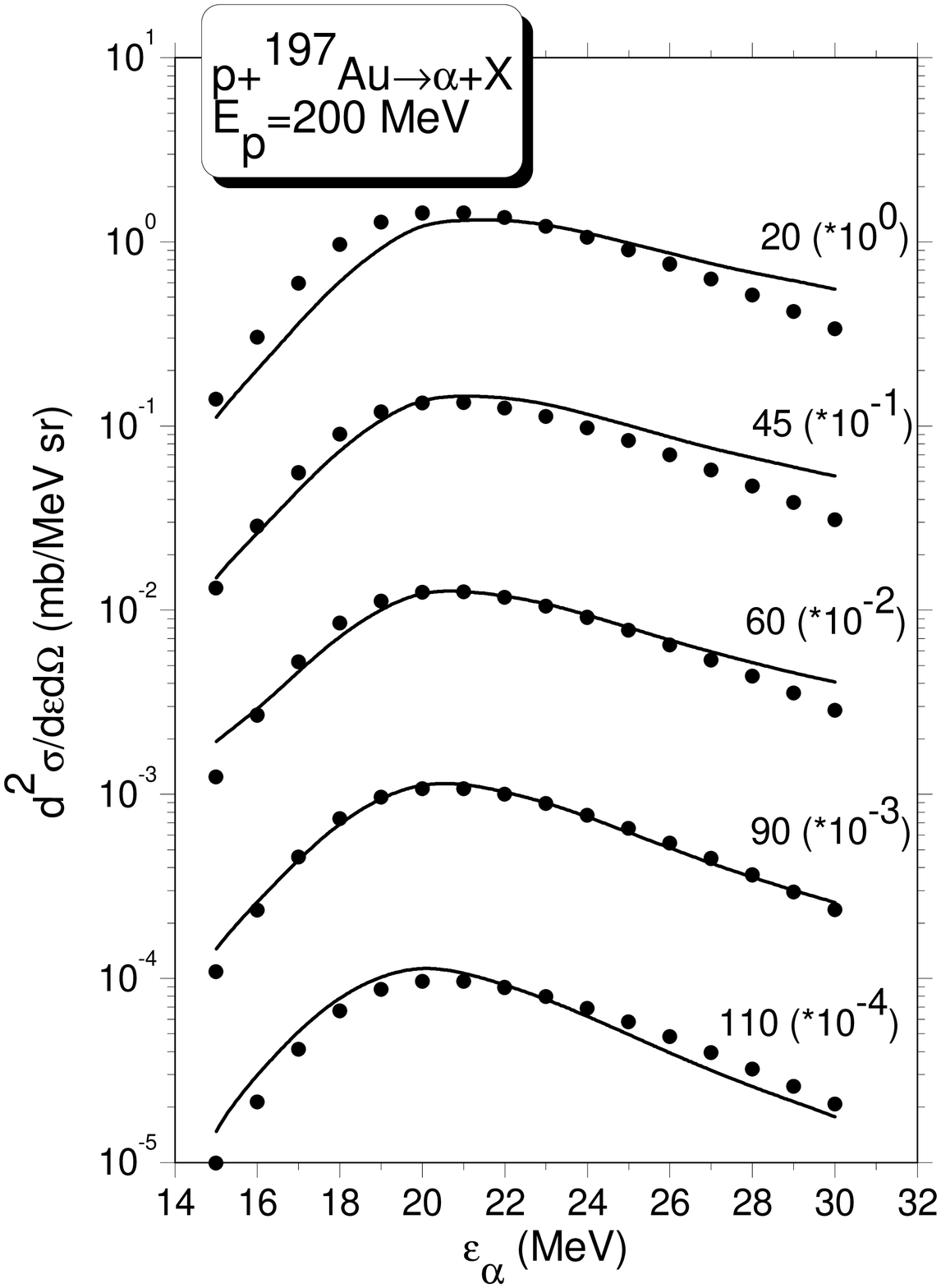}
\caption{Fit with one source to the low energy data. The laboratory
emission angle is given next to the appropriate data set or fit
curve, respectively. The data and the fits were multiplied with the
indicated factors.} \label{fig:Moving_source_fit}
\end{figure}

In order to test the assumption that the low energy part is
dominated by barrier penetration we have fitted one single source
with a constant temperature and a constant source velocity to the
cross sections in the range from 15--30 MeV. The usual practice, as
we have also applied above, namely, to correct the energy by
subtracting the Coulomb barrier energy, is not applicable since it
leads to negative energies. Consequently, Eq.
(\ref{eqn:moving_source}) can not reproduce the data. In this case
one has to take the barrier penetration explicitly into account. We
multiply the r.h.s. of Eq. (\ref{eqn:moving_source}) by the
penetration probability \cite{Wong72, Wong73, Machner85}
\begin{equation}
P(\epsilon)=\frac{\hbar \omega}{2\pi\epsilon}\ln \left\{1+ \exp
\left[ \frac{2\pi\left( \epsilon -V_C\right)}{\hbar
\omega}\right]\right\}
\end{equation}
where $\omega$ is the frequency associated with a mean potential to
be tunnelled through. Whereas fitting to an excitation function
$\alpha + ^{238}U\to$ fission \cite{Viola62} leads to $\hbar
\omega\approx 4$ MeV \cite{Wong73}, the present result is 6.0(3)
MeV. The rather small Coulomb barrier of 17.67(23) MeV corresponds
to a large radius of the emitting system. This might be an
indication that the highly excited nucleus has expanded. For the
source velocity the fit results in 0.0025(4)$c$, while one would
expect 0.0033$c$ from momentum conservation. This is an indication
of fast particle emission in the equilibration process. The result
of this exercise is shown in Fig. \ref{fig:Moving_source_fit}.
Finally we report a fit value for the temperature of 3.12(17) MeV.
There might be a correlation between the fit parameters. We have
therefore performed moving source fits with barrier penetration to
the angle integrated spectra, which will be discussed below.
Unfortunately the different components are not clearly
distinguishable as they are for the case of $\alpha$-particle
emission from gold, thus resulting in fits which are not so good.
But again we find rather small values for the barrier. It will be
interesting to study further data around the barrier and to see
whether the barrier is reduced in comparison to a nucleus in its
ground state.

We use the slope and intercept parameters in Eq. (\ref{eqn:slope})
to get angle integrated cross sections. The angle integrated cross
section is
\begin{equation}
\frac{d\sigma(\epsilon)}{d\epsilon} = \frac{2\pi}{a(\epsilon)}
\left\{\exp\left[b(\epsilon)+a(\epsilon)\right] -
\exp\left[b(\epsilon)-a(\epsilon)\right]\right\}.
\end{equation}
The resulting differential cross sections for the three targets are
shown in Figs \ref{fig:int_al}--\ref{fig:int_au}. For the two
lighter targets the energy distributions show an almost exponential
slope without structure. In case of the gold target this structure
is modified due to Coulomb effects. The distributions are discussed
below.

\section{Data Comparison}\label{sec:Data-Comparison}
\begin{figure}[h]
\begin{center}
\includegraphics[width=7 cm]{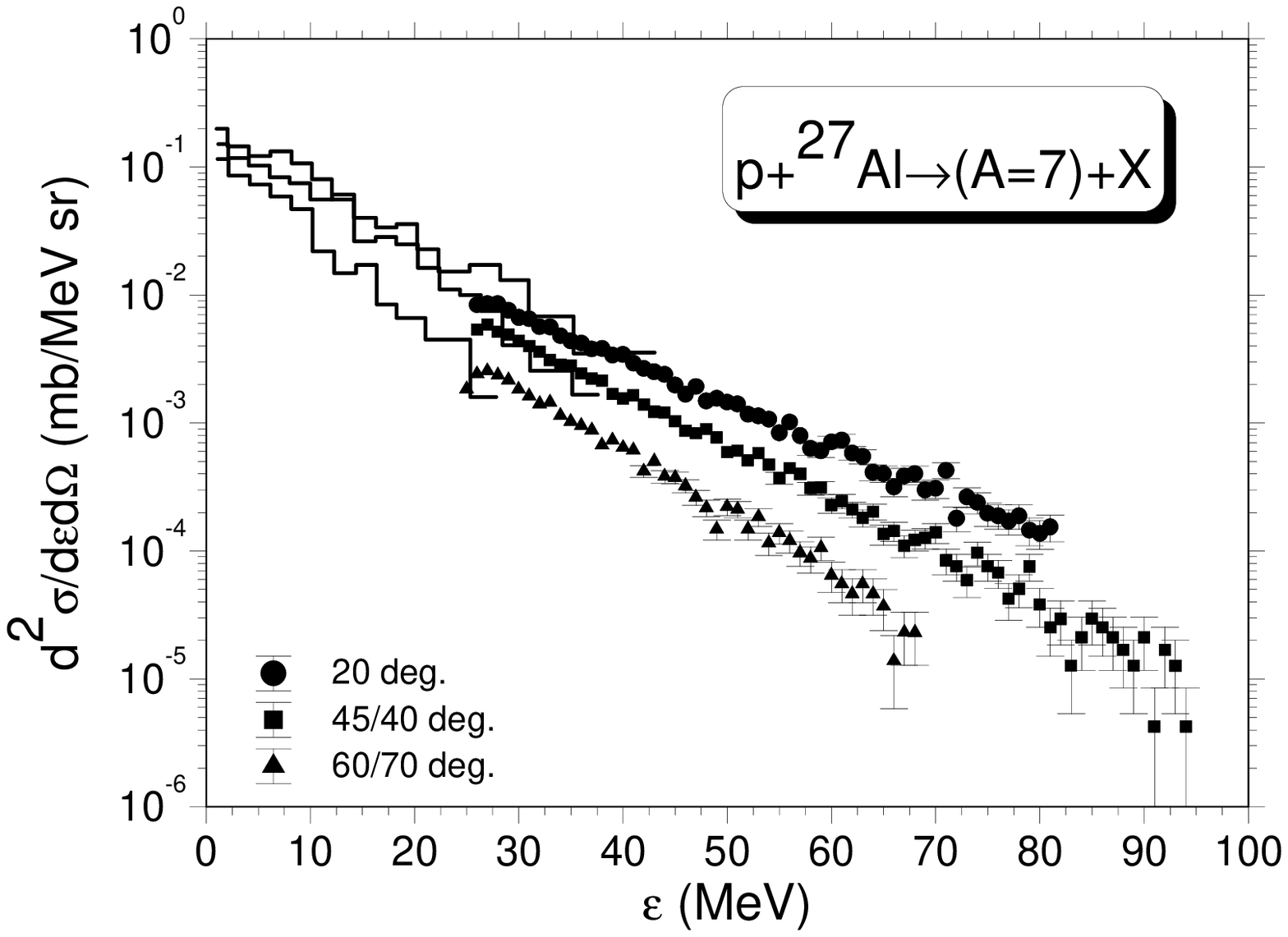}
\caption{Double differential cross sections for $A=7$ emission
following the bombardment of $^{27}Al$ with protons. The data from
Ref. \cite{Kwiatkowski83} are shown as histograms and are for 180
MeV, the present data are shown as dots with error bars. The
emission angles are given in the legend where the first angle is for
the present data, the second for those from Ref.
\cite{Kwiatkowski83}.} \label{Fig:comparison_NAC_Indiana}
\end{center}
\end{figure}

Although there are no data on IMF cross sections with exactly the
same beam energy and for the same targets as employed in this study,
there are data for energies or targets close by. We will compare the
present data with those. First we will compare differential cross
sections for the reaction $p+^{27}Al\to (A=7)+X$. For that purpose
the present cross sections for $^7Li$ and $^7Be$ emission were
added. In Fig. \ref{Fig:comparison_NAC_Indiana} these cross sections
are compared with those from Kwiatkowski et al. \cite{Kwiatkowski83}
taken at a beam energy of 180 MeV. They measured fragments with
masses $A\ge 6$ and energies down to $\epsilon/A\ge 0.05$ MeV/$u$.
The data have only a moderate overlap with the present data. There
seems to be consistency between both data sets with respect to the
absolute height as well as the shape of the spectra.

As already stated in the introduction there are two studies of IMF
emission from $Ag$. One was performed at a proton beam energy of 161
MeV \cite{Yennello90}. Also these data cover smaller fragment
energies than the present due to a gaseous $\Delta E$ detector. They
observed fragments with charge number $Z$ ranging from 3 to 12.
\begin{figure}[h]
\begin{center}
\includegraphics[width=7 cm]{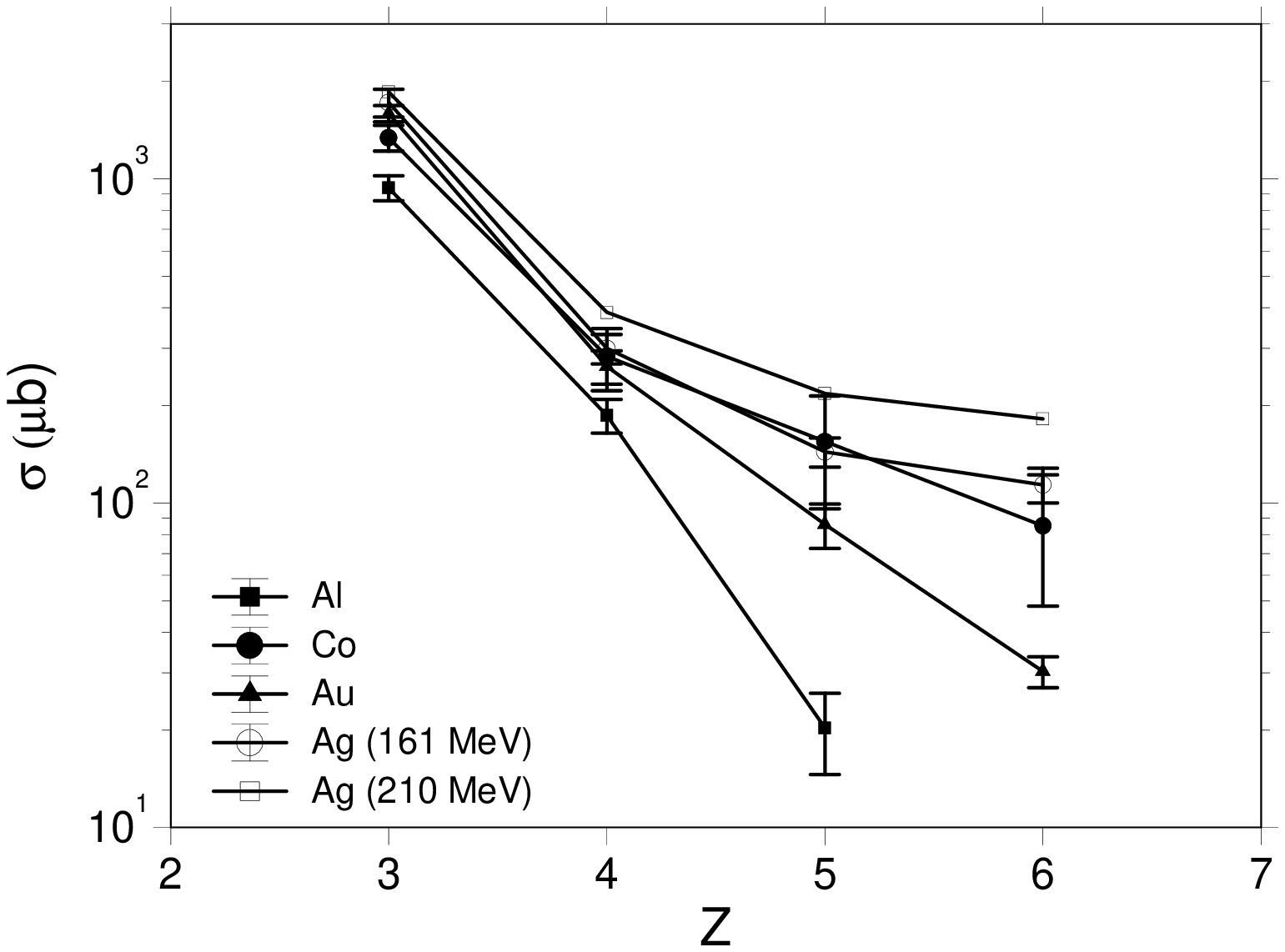}
\caption{Cross sections for total IMF production. The present data
are shown by full symbols, those from Ref. \cite{Yennello90} by open
dots and those by Ref. \cite{Green80} by open squares.}
\label{Fig:Z-distribution}
\end{center}
\end{figure}
The total cross sections from this measurement are shown in Fig.
\ref{Fig:Z-distribution} as function of the fragment charge number
together with those from Green and Korteling \cite{Green80} and the
present results. For the case of the aluminum targets a large
fraction of the cross section is missing due to the thickness of the
first $\Delta E$ detector used here (see the latter comparison with
model calculations and Figs. \ref{fig:int_al}-\ref{fig:int_au}).
This makes the discrepancy in the yields esp. for $Z=5$ and 6. The
yield in the case of the cobalt target agrees best in that of the
silver target.
\begin{figure}
\begin{center}
\includegraphics[width=7 cm]{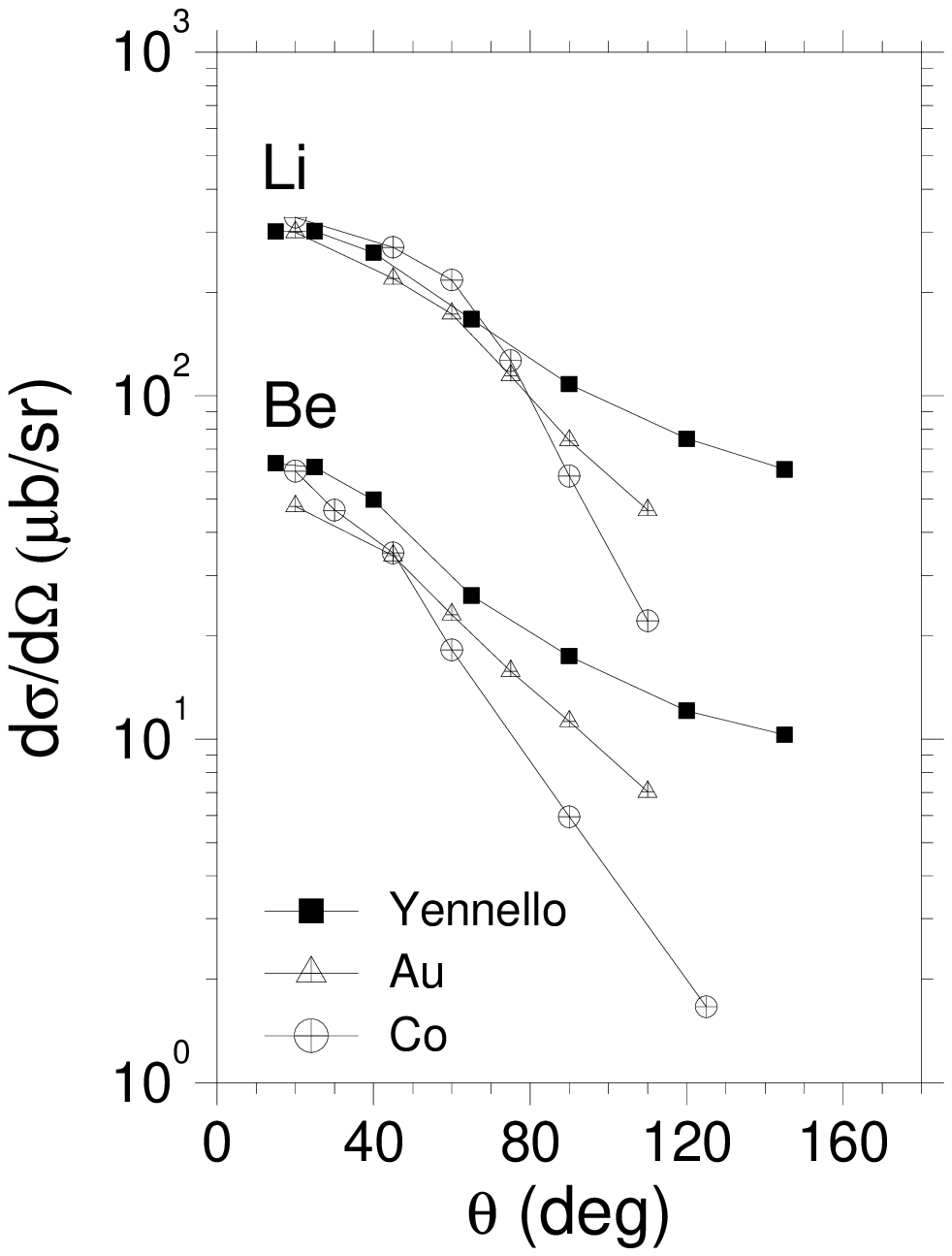}
\caption{Angular distributions of $Li$ and $Be$ fragments for the
indicated targets (this work) and for a silver target at 161 MeV
proton energy \cite{Yennello90}.} \label{Fig:Comp_MD}
\end{center}
\end{figure}
In Fig. \ref{Fig:Comp_MD} angular distributions for $Li$ and $Be$
fragments integrated over the acceptance range are compared with
those of Ref. \cite{Yennello90}. Again the agreement is reasonable
with respect to the different energy ranges in the different
experiments. Summarizing this comparison one can state that there is
a fair agreement between the different measurements for $Z=3$ and 4.
It may be more instructive to continue the comparison on the level
of spectra.

\begin{figure}
\begin{center}
\includegraphics[width=7 cm]{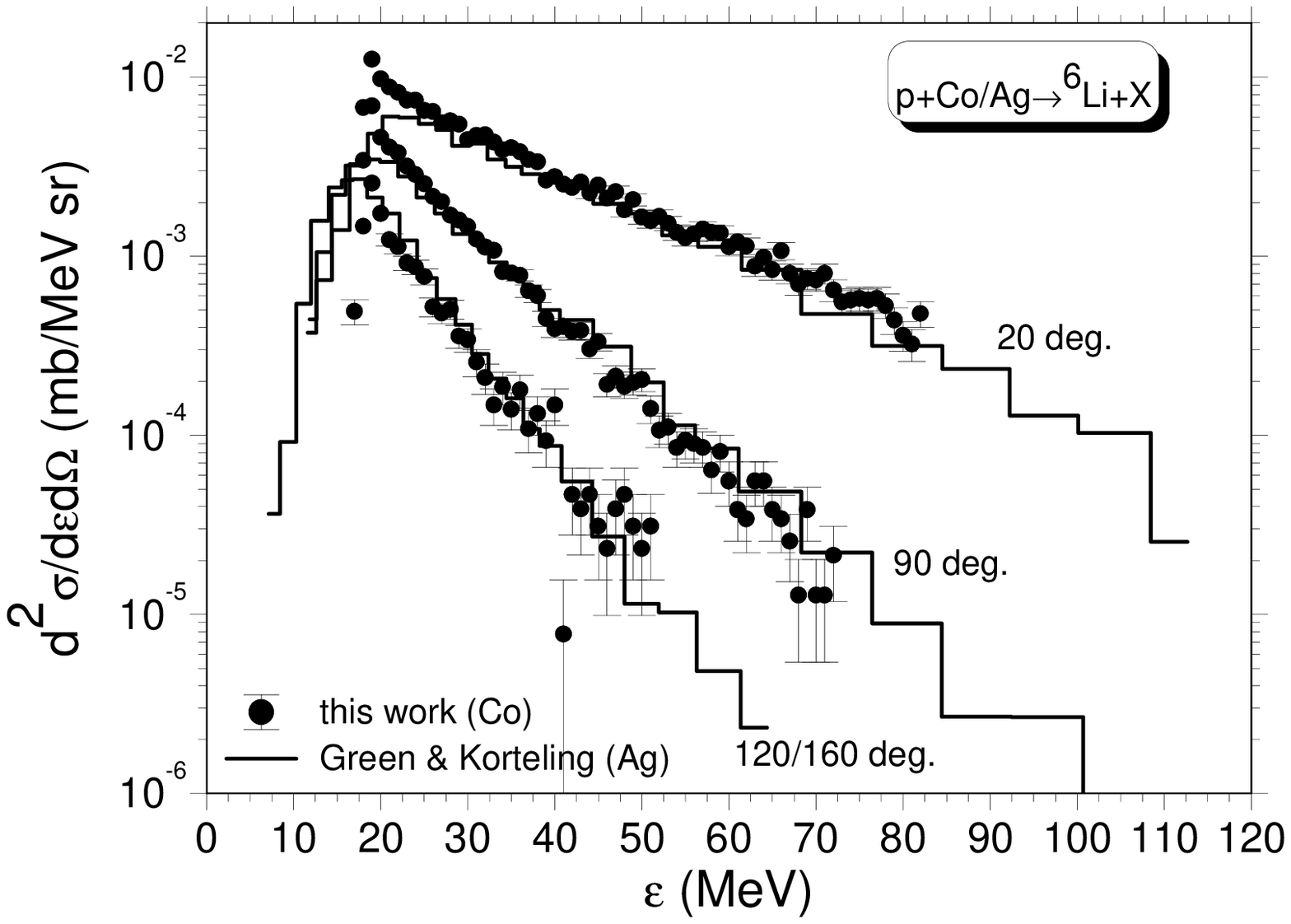}
\caption{Comparison of double differential cross sections from the
$p+^{59}Co\to ^6Li+X$ reaction (this work, dots with error bars) and
$p+Ag\to ^6Li+X$ reaction (Ref. \cite{Green80}, histograms). The
first angle given is for the present data while the second is for
those from \cite{Green80}.} \label{Fig:compare_NAC_TRIUMF}
\end{center}
\end{figure}
This is done in Fig. \ref{Fig:compare_NAC_TRIUMF} for the case of
isotopic $^6Li$ emission from $Ag$ (210 MeV, Ref. \cite{Green80})
and $^{59}Co$. There is excellent agreement between the two data
sets  with respect to shape and absolute height. Finally we compare
the slope of the present data in case of $\alpha$-particle emission
for the aluminum and gold target with those of ref.
\cite{Didelez82}. The later have been multiplied with an overall
normalization factor of four. It becomes clear that the angular
dependencies agree with each other in the overlap region.
\begin{figure}
\includegraphics[width=7 cm]{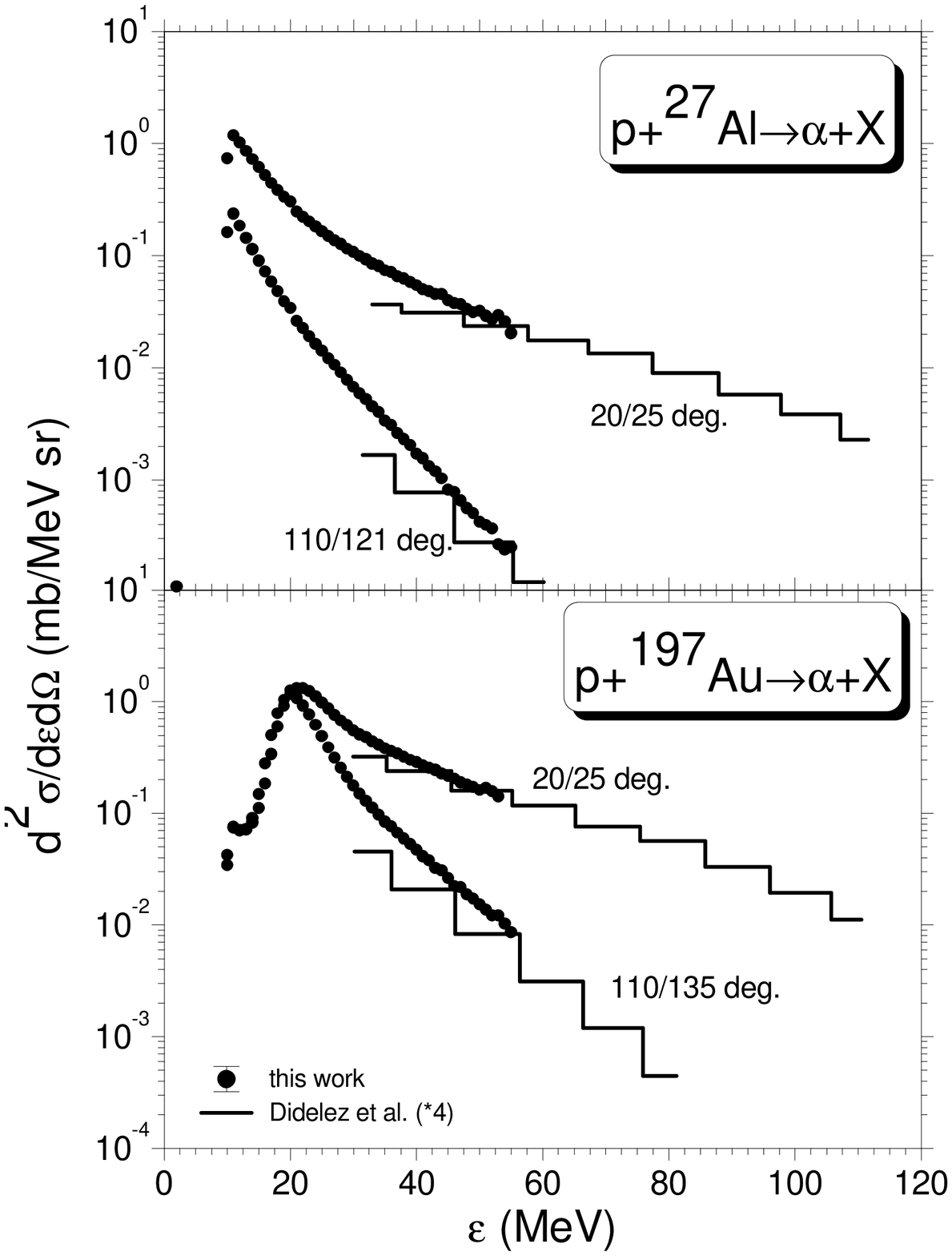}
\caption{Similar as Fig. \ref{Fig:compare_NAC_TRIUMF} but for the
indicated reactions and data from Ref. \cite{Didelez82}, normalized
by a factor of four.} \label{Fig:comp_NAC_ORSAY}
\end{figure}

From these comparison it becomes evident that one can state that the
present data are correct with respect to spectral shapes, angular
distributions and absolute magnitude of the cross sections.
\newpage
\section{Comparison with model calculations}
In this section model calculations are compared to the deduced data.
A variety of models for fast particle emission in nuclear reactions
are discussed in Ref. \cite{Machner85}. A model especially suited
for higher projectile energies is the intranuclear cascade model
(INC). Although it cannot account for IMFs during the equilibration
process it predicts the final excitation energy of an equilibrated
system. This system then undergoes de-excitation by evaporating
lower energy particles.

The INC model was first proposed by Serber in 1947 \cite{Serber47}.
The successful realization of this model by means of Monte Carlo
simulations was published by Goldberger who did the first
calculations by hand in 1948 \cite{Goldberger48}. Computer
simulations were first done by Metropolis et al.
\cite{Metropolis58}. In the present work we have applied the model
in the standard Li$\grave{e}$ge version INCL4.2 \cite{Cugnon87}.

The Intranuclear Cascade (INC) Model simulates - by the Monte Carlo
method - sequences of the nucleon-nucleon collisions proceeding
inside the nucleus.   This is equivalent to solving the Boltzmann
transport equation for the time dependent distribution of the
nucleons in the nucleus, treating explicitly collisions between the
nucleons. As mentioned above, such a picture of the reaction is
justified in the case when the energy of the projectile is high
enough. The INC is stopped when signatures are fulfilled, which
indicate equilibration of the decaying nucleus. In the INCL4.2 code
the equilibration time $\tau$ is determined by reaching a constant
emission rate of cascade particles during the INC process. Typically
$\tau$ is of the order of $10^{-22}$s or 30 fm/c. The longer this
somewhat ``artificially'' chosen time the smaller $E^*$ is left for
the evaporation process. Here we have chosen
\begin{equation}
\tau = \tau_0 \left(\frac{A_T}{208}\right)^{0.16}
\end{equation}
with $\tau_0=44.1$ fm/c. This is smaller than the value used in
\cite{Boudard02} but corresponds to the earlier used value
\cite{Cugnon97}.

The description of each cascade involves three different stages: (i)
initialization of the properties of the spatial and momentum
distribution of nucleons in the nucleus, (ii) propagation of
nucleons inside the nucleus, and (iii) collisions of the nucleons.

The simplicity of the model and speed of calculations makes the INCL
model very attractive. Of course the model cannot efficiently
describe evaporation of the particles from the compound nucleus
formed in the first stage of the reaction for two reasons: (a) the
evaporation is very sensitive to the density of states of the nuclei
participating in the reaction, whereas the single particle density
of states implicitly present in the INCL calculations is not exact
enough, and (b) the calculations of the cascade over such long times
as those characteristic for the compound nucleus emissions are not
stable numerically and very inefficient. The other very serious
drawback of the INCL model is absence of correlations between
nucleons, which could lead to emission of complex fragments.   This
is because the INCL is a single particle model with the mean field
treated in an oversimplified manner.  The mean field used in the
INCL makes the assumption of being constant throughout the volume of
the nucleus or modified at the surface of the nucleus, but it is
always a static field.

In practice the model of the intranuclear cascade (or equivalent) is
applied to describe the first stage of the nuclear collision and the
calculations are stopped once it can be assumed that equilibrium has
been achieved.  In the present study the INCL4.2 computer code was
applied for this purpose. Discussion concerning the criteria for the
terminating of the intranuclear cascade are presented by J. Cugnon
et al. in Ref. \cite{Cugnon97}.

After equilibration is reached we apply an evaporation model. It is
the generalized evaporation model (GEM) of Furihata
\cite{Furihata00}, which is based on the classical Weisskopf --
Ewing approach \cite{Weisskopf37,Weisskopf40}.  According to this
approach, the probability of evaporation of the particle $j$ from a
parent compound nucleus $i$ with a total kinetic energy in the
center-of-mass system between $\epsilon$ and $\epsilon$+d$\epsilon$
is defined as:
\begin{equation}
P_{j}(\epsilon)d\epsilon=g_{j}\sigma_{inv}(\epsilon)\frac{\rho_{d}(E-Q-\epsilon)}{\rho_{i}(E)}\epsilon
d\epsilon, \label{prob}
\end{equation}
where $E$ is the excitation energy of the parent nucleus $i$, $d$
denotes a daughter nucleus produced after the emission of ejectile
$j$, and $\rho_i$, $\rho_d$ are the level densities for the parent
and daughter nucleus respectively.
 $Q$ denotes the $Q$ -- value of the reaction. The statistical and
normalization factor $g_{j}$ is defined as \begin{math}
g_{j}=(2S_{j}+1)m_{j}/\pi^{2}\hbar^{2} \end{math}, where $S_{j}$ and
$m_{j}$ are the spin and the mass of the emitted particle $j$
respectively. The cross section $\sigma_{inv}$ for the inverse
reaction is evaluated from
\begin{equation}
\sigma_{inv}(\epsilon)=\sigma_{g}P(\epsilon)) \label{equ:cross}
\end{equation}
where  $\sigma_{g}$ is the geometrical cross section. GEM considers
fragments heavier than helium nuclei. There are 66 ejectiles (see
Table \ref{tab:ejectiles}).  For the barrier penetration probability
$P$ we have used the form of Ref. \cite{Dostrovsky59}. The
parameters for light particles (n, p, d, t, $^3$He and $^4$He) are
taken from Ref. \cite{Dostrovsky59} whereas those for IMFs were
adopted from the work of Matsuse et al. \cite{Matsuse82}.

\begin{table}
\caption{\label{tab:ejectiles} The ejectiles taken into
consideration in the GEM calculations.}
\begin{center}
\begin{tabular}{rlllllll}
\hline \hline
$Z_j$  & Ejectiles &           &           &           &           &           &  \\
\hline
0      &   n       &           &           &           &           &           &  \\
1      &   p       & d         &      t    &           &           &           &  \\
2      & $^3$He    & $^4$He    & $^6$He    & $^8$He    &           &           &  \\
3      & $^6$Li    & $^7$Li    & $^8$Li    & $^9$Li    &           &           &  \\
4      & $^7$Be    & $^9$Be    & $^{10}$Be & $^{11}$Be & $^{12}$Be &           &  \\
5      & $^8$B     & $^{10}$B  & $^{11}$B  & $^{12}$B  & $^{13}$B  &           &  \\
6      & $^{10}$C  & $^{11}$C  & $^{12}$C  & $^{13}$C  & $^{14}$C  & $^{15}$C  & $^{16}$C \\
7      & $^{12}$N  & $^{13}$N  & $^{14}$N  & $^{15}$N  & $^{16}$N  & $^{17}$N  &          \\
8      & $^{14}$O  & $^{15}$O  & $^{16}$O  & $^{17}$O  & $^{18}$O  & $^{19}$O  & $^{20}$O \\
9      & $^{17}$F  & $^{18}$F  & $^{19}$F  & $^{20}$F  & $^{21}$F  &           &          \\
10     & $^{18}$Ne & $^{19}$Ne & $^{20}$Ne & $^{21}$Ne & $^{22}$Ne & $^{23}$Ne & $^{24}$Ne \\
11     & $^{21}$Na & $^{22}$Na & $^{23}$Na & $^{24}$Na & $^{25}$Na &           &           \\
12     & $^{22}$Mg & $^{23}$Mg & $^{24}$Mg & $^{25}$Mg & $^{26}$Mg & $^{27}$Mg & $^{28}$Mg \\
\hline \hline
\end{tabular}
\end{center}
\end{table}

The total decay width $\Gamma_{j}$ is calculated by integrating Eq.
(\ref{prob}) using  Eq. (\ref{equ:cross}) and is expressed as:
\begin{equation}
\label{eq:Gamma}
\Gamma_{j}=\frac{g_{j}\sigma_{g}}{\rho_{i}(E)}\int_{V}^{E-Q}\epsilon
P(\epsilon)\rho_{d}(E-Q-\epsilon)d\epsilon.
\end{equation}

where $V$ is the Coulomb barrier. For the level density we have
applied the  Fermi gas model expression
\begin{eqnarray}
\rho (E) & = &\frac{\pi}{12}\frac{e^{2\sqrt{a(E-\delta)}}}{a^{1/4}(E-\delta)^{5/4}} \;\;\;\textrm{for}~E\geq E_{x} \\
 & = &\frac{\pi}{12}\frac{1}{T}e^{(E-E_{o})/T} \;\;\; \;\;\;\;\;\; \textrm{for}~E\leq E_{x}
\end{eqnarray}
where $a=A_d/8$ (MeV$^{-1}$) is the level density parameter,
$\delta$ is the pairing energy of the residual, $T$ is again the
nuclear temperature given by $1/T=\sqrt{a/U_{x}}-1.5/U_{x}$ where
$U_{x}$ is defined as $U_{x}=2.5+150/A_d$. The excitation energy
E$_x$, for which the formula for level density changes its form, is
evaluated as E$_x$=U$_x$+$\delta$.  To obtain a smooth continuity
between the two formulae, the E$_0$ parameter is determined as
follows:
\begin{equation}
E_0 = E_x -T \cdot (\ln T -0.25 \ln a - 1.25 \ln U_x +2 \sqrt{a
U_x}).
\end{equation}
The contribution of the emission of IMFs in a long living excited
state is taken into account together with those which decay to the
ground state. The condition for the lifetime of excited nuclei
considered in GEM is as follows:
\begin{math}   T_{1/2}/ln2 > \hbar/\Gamma_{j}^{*} \end{math}.
The value of $\Gamma_{j}^{*}$ is defined as the emission width of
the decaying ejectile and is calculated in the same way as for the
ground state, i.e. by Eq. (\ref{eq:Gamma}). The total emission width
of an ejectile is summed over its ground state and all its excited
states. All input parameters are the standard parameters of the
models. We have not adjusted parameters to fit the present data.

\begin{figure}
\begin{center}
\includegraphics[width=7 cm]{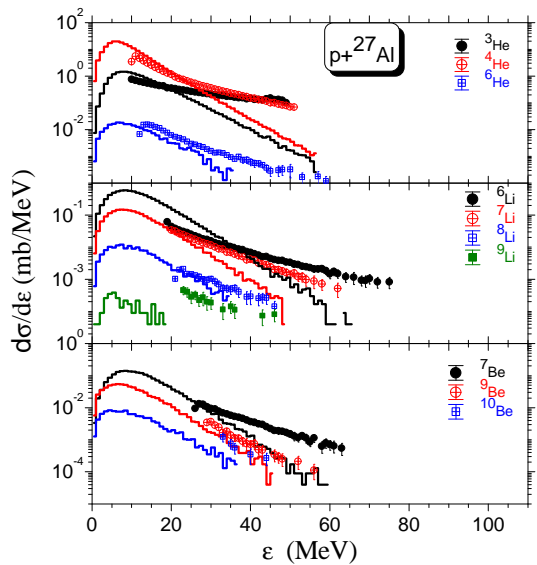}
\caption{(Color online) Energy differential cross sections for
indicated IMFs for aluminum. The data are shown by the different
symbols indicated in the figure. The histograms are calculations
described in the text. } \label{fig:int_al}
\end{center}
\end{figure}
\begin{figure}
\begin{center}
\includegraphics[width=7 cm]{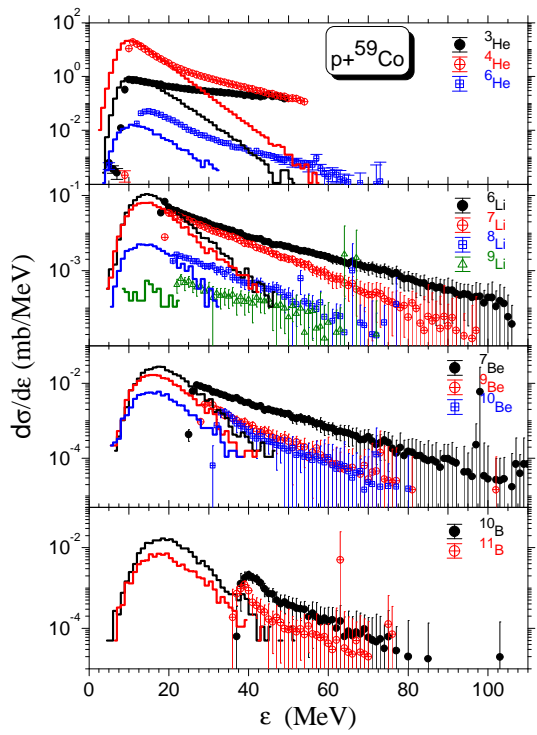}
\caption{(Color online) Same as Fig. \ref{fig:int_al} but for
cobalt.} \label{fig:int_co}
\end{center}
\end{figure}
\begin{figure}
\begin{center}
\includegraphics[width=7 cm]{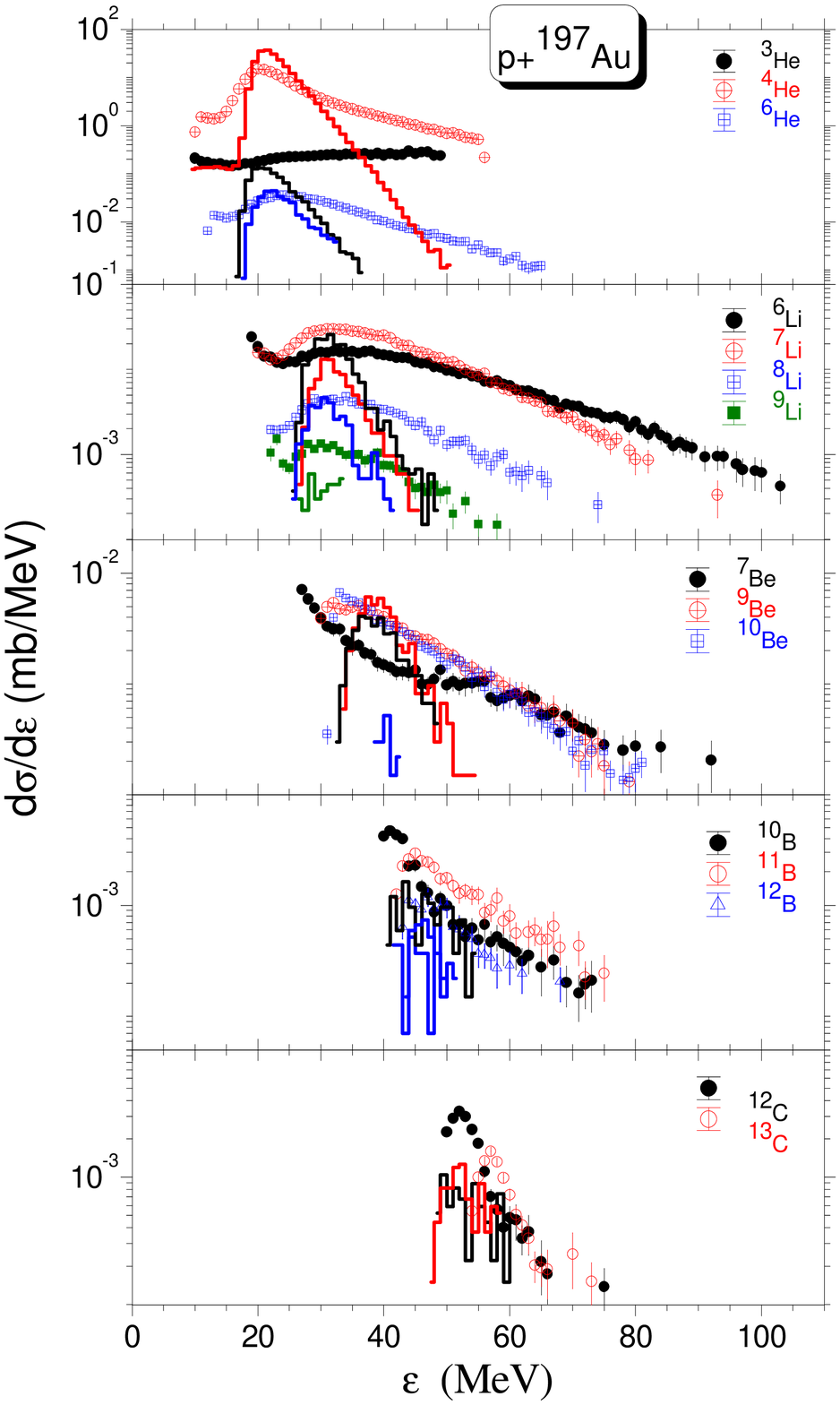}
\caption{(Color online) Same as Fig. \ref{fig:int_al} but for gold.}
\label{fig:int_au}
\end{center}
\end{figure}

In Figs \ref{fig:int_al}--\ref{fig:int_au} we compare the angle
integrated cross sections with the results of the calculations
sketched above.  A general trend is visible. For high IMF energies
the calculation underestimates the experiment. The non-equilibrium
fraction of the cross section is quite large in agreement with other
experiments \cite{Green80, Yennello90}. The heavier the IMFs are,
the more the agreement between the calculations and data
deteriorates, even in the evaporative region. Heavy IMFs are
strongly underestimated. In the case of gold there is emission of
IMFs from lighter composite systems, most probably fission
fragments. This is also visible in the calculation but to a lesser
extent than is observed in the data. The sequence within the
isotopes is always obeyed by the calculations. However, in the case
of $^7$Be emission from a gold-like composite the calculation
predicts an almost negligible cross section, while the experiment is
orders of magnitude higher. This is especially true for low energies
and may point again to emission from lighter sources than the
target-like system.

\section{Discussion}
We have measured IMF (He--C) emission for $^{27}$Al, $^{59}$Co and
$^{197}$Au targets at a proton beam energy of 200 MeV, which is near
the maximal abundance in the proton distribution in cosmic rays. The
fragments were isotopically resolved. Spectra were taken at
laboratory angles from 20\Grad to 120\Grad. Analysis in terms of a
model of a moving source with continuous temperature and source
velocity shows a linear relationship between these two quantities as
a function of particle energy. The data in the case of gold show a
strong influence of the Coulomb barrier. In the cases of the two
lighter targets this feature was suppressed by the thickness of the
first $\Delta E$ counter. Emission of fragments with a significantly
smaller Coulomb barrier than for a target-like system is observed.
The assumption that we observe emission from excited fission
fragments was studied in evaporation calculations. Indeed, the
calculations also show such fragments (see $\alpha$-particle
emission in Fig. \ref{fig:int_au}), although with a cross section
more than one order of magnitude smaller than the experiment. The
data were compared to model calculations. The first stage was
calculated with an intranuclear cascade. In this cascade the
emission of pions and nucleons can take place. After equilibrium is
reached the energy and momentum distribution of the excited
composite is transferred to an evaporation model, which, in addition
to nucleons, allows for IMF emission. The frequency of isotopes
being emitted for a specific element is followed by the calculation.
The high energy tails visible in the experiments are not reproduced
by the calculations. The emission of the heavier isotopes like
$^9$Li, $^{10}$Be, boron and carbon but also $^7$Be is
underestimated.

The reduction of the Coulomb barrier observed may be due to dilution
of the composite system. But an effect originating from the fission
fragments can not be excluded. For the other two target nuclei we
could not measure below the Coulomb barrier. Such data are highly
desired to answer this question. Calculations, treating also IMF
emission in the first fast stage are not yet available for the
present data but are also desired.

Let us now come back to the problem of Galactic cosmic rays (GCR) as
discussed in the introduction. The production of $BeB$ by Galactic
cosmic-ray  spallation of interstellar $CNO$ nuclei was the standard
model for $BeB$ nucleosynthesis for almost two decades after first
being proposed \cite{Reeves70, Meneguzzi71}. However, this simple
model was challenged by the observations of $BeB$ abundances in
Population II stars, and particularly the $BeB$ trends versus
metallicity. Measurements showed that both $Be$ and $B$ vary roughly
linearly with $Fe$, a "primary" scaling. In contrast, standard GCR
nucleosynthesis predicts that $BeB$ should be "secondary" versus
spallation targets $CNO$, giving $Be\propto O^2$
\cite{Vangioni-Flam90}. If $O$ and $Fe$ are coproduced (i.e., if the
ratio $O/Fe$ is constant), then the data clearly contradict the
canonical theory, i.e., $BeB$ production via standard GCR's
\cite{Fields00}. In order to accurately calculate the effects of the
propagation of cosmic ray nuclei in the galaxy one needs to
incorporate at least several hundred secondary cross sections into
the propagation calculation. For charges with $Z<28$ this involves
the fragmentation from $\approx 55$ nuclei with mass numbers A
between 6 and 60 \cite{Webber98}. The present data should help to
improve our understanding of the systematics of the cross sections
as a function of Z, A, and A/Z.

In GCR's one observes only stable isotopes since short lived
isotopes decay. Thus only $^3He$ is observed because all tritium
decays into it. The only difference might be $^{10}Be$ which has a
half life of $1.6\times 10^6$ years. We therefore compare the ratio
between $^{10}Be$ and $^{9}Be$ for the different targets.  The yield
of the short lived $^{9}Li$ is added to the latter. We report the
ratio of the cross sections integrated over an energy range from 30
to 50 MeV in Table \ref{Tab:Ratio}. Within this range for all three
particle types data exist.
\begin{table}
\centering \caption{The ratio $R=\sigma(^{10}Be) / [\sigma(^{9}Be) +
\sigma(^{9}Li)]$ for the different targets in the energy range 30 to
50 MeV.}\label{Tab:Ratio}
\begin{tabular}{l|c}
\hline \hline
target & R \\
\hline
$^{27}Al$ & $0.654\pm 0.073$ \\
$^{59}Co$ & $0.577\pm 0.133$ \\
$^{197}Au$ & $0.918\pm 0.109$ \\
\hline \hline
\end{tabular}
\end{table}
The ratio in cases of the two lighter targets is within error bars
identical. For gold the primordial abundance of $^{10}Be$ relative
to the $A=9$ fragments is much larger than in the case of the two
lighter targets. These ratios should be essential to study the age
of GCR's.

\section{Acknowledgement}
We acknowledge the cyclotron crew for providing us with the
excellent beam. We thank C. J. Stevens (mechanics) and V. C. Wikner
(electronics) for technical help in the preparation of the
experiment.


\begin{thebibliography}{10}

\bibitem{Auger39}
P.~Auger, R.~Maze, P.~Ehrenfest, A.~Fr\'{e}on: J. Phys. Radium, {\bf
10} (1939)
  39.

\bibitem{Brown49}
R.H. Brown, et~al.: Philos. Mag., {\bf 40} (1949) 862.

\bibitem{Powell59}
C.F. Powell, P.H. Fowler, D.H. Perkins:  (1959) The Study of
Elementary
  Particles by the Photographic Method, Pergamon Press, New York.

\bibitem{Fields00}
B.~D. Fields, K.~A. Olive, E.~Vangioni-Flam, M.~Cass\'{e}:
Astrophysical
  Journal, {\bf 540} (2000) 930.

\bibitem{Biermann01}
P.~L. Biermann, N.~Langer, Eun-Suk Seo, T.~Stanev: Astronomy $\&$
Astrophysics,
  {\bf 369} (2001) 269.

\bibitem{Hoerandel01}
J.~R. H\"{o}randel, et~al.:  (2001) in Proceedings of 27th
International Cosmic
  Ray Conference, Copernicus Gesellschaft (www.copernicus.org/C4/index.htm),
  page 1608.

\bibitem{Nolfo01}
G.~A. de~Nolfo~et al.:  (2001) in Proceedings of 27th International
Cosmic Ray
  Conference, Copernicus Gesellschaft (www.copernicus.org/C4/index.htm), page
  1667.

\bibitem{Michel96}
R.~Michel, I.~Leya, L.~Borges: Nucl. Instruments and Meth. in Phys.
Res., {\bf
  B 113} (1996) 343.

\bibitem{Waddington99}
C.J. Waddington, J.R. Cummings, B.S. Nilsen, T.L. Garrard: Astr.
Phys. J., {\bf
  519} (1999) 214.

\bibitem{Huefner85}
J.~H\"{u}fner: Phys. Rep., {\bf 125} (1985) 129.

\bibitem{Silberberg90}
R.~Silberberg, C.H. Tsao: Phys. Rep., {\bf 191} (1990) 351.

\bibitem{Austin81}
S.~Austin: Progress in Part. Nucl. Phys., {\bf 7} (1981) 1.

\bibitem{Simpson83}
J.~A. Simpson: Ann. Rev. Nucl. Part. Sci., {\bf 33} (1983) 323.

\bibitem{Didelez82}
J.~P.~Didelez et~al.: www.fz-juelich.de/ikp/gem under Conference
Proceedings
  and Ricerca Scientifica ed Educazione Permanente, {\bf Suppl. 28} (1982) 237.

\bibitem{Machner84}
H.~Machner et~al.: Phys. Lett., {\bf B 138} (1984) 39.

\bibitem{Kalbach88}
C.~Kalbach: Phys. Rev., {\bf C 37} (1988) 2350.

\bibitem{Kalbach87}
C.~Kalbach: priv. communication to H. M.

\bibitem{Green80}
R.~E.~L. Green, R.~G. Korteling: Phys. Rev., {\bf C 22} (1980) 1594.

\bibitem{Yennello90}
S.~J.~Yennello et~al.: Phys. Rev., {\bf C 41} (1990) 79.

\bibitem{Pilcher89}
J.~V. Pilcher, A.~A. Cowley, D.~M. Whittal, J.~J. Lawrie: Phys.
Rev., {\bf C
  40} (1989) 1973.

\bibitem{Markiel88}
W.~Markiel et~al.: Nucl. Phys., {\bf A 485} (1988) 445.

\bibitem{Machner90}
H.~Machner: Z. Physik, {\bf A 336} (1990) 209.

\bibitem{Wong72}
C.~Y. Wong: Phys. Lett., {\bf B 42} (1972) 156.

\bibitem{Wong73}
C.~Y. Wong: Phys. Rev. Lett., {\bf 31} (1973) 77.

\bibitem{Machner85}
H.~Machner: Phys. Rep., {\bf 127} (1985) 309.

\bibitem{Viola62}
V.~E. Viola, T.~Sikkeland: Phys. Rev., {\bf 128} (1962) 767.

\bibitem{Kwiatkowski83}
H.~Kwiatkowski et~al.: Phys. Rev. Lett., {\bf 50} (1983) 1648.

\bibitem{Serber47}
R.~Serber: Phys. Rev., {\bf 72} (1947) 1113.

\bibitem{Goldberger48}
M.~Goldberger: Phys. Rev., {\bf 74} (1948) 1269.

\bibitem{Metropolis58}
N.~Metropolis et~al.: Phys. Rev., {\bf 110} (1958) 185.

\bibitem{Cugnon87}
J.~Cugnon: Nucl. Phys., {\bf A 462} (1987) 751.

\bibitem{Boudard02}
A.~Boudard, J.~Cugnon, S.~Leray, C.~Volant: Phys. Rev., {\bf C 66}
(2002)
  044615.

\bibitem{Cugnon97}
J.~Cugnon, C.~Volant, S.~Vuillier: Nucl. Phys., {\bf A 620} (1997)
475.

\bibitem{Furihata00}
S.~Furihata: Nucl. Instruments and Meth. in Phys. Res., {\bf B 171}
(2000) 251.

\bibitem{Weisskopf37}
V.~F.~Weisskopf et~al.: Phys. Rev., {\bf 52} (1937) 295.

\bibitem{Weisskopf40}
V.~F. Weisskopf, D.~H. Ewing: Phys. Rev., {\bf 57} (1940) 472.

\bibitem{Dostrovsky59}
I.~Dostrovsky et~al.: Phys. Rev., {\bf 116} (1959) 683.

\bibitem{Matsuse82}
T.~Matsuse, A.~Arima, S.M. Lee: Phys. Rev., {\bf C 26} (1982) 2338.

\bibitem{Reeves70}
H.~Reeves, W.~A. Fowler, F.~Hoyle: Nature, {\bf 226} (1970) 727.

\bibitem{Meneguzzi71}
M.~Meneguzzi, J.~Audouze, H.~Reeves: Astronomy $\&$ Astrophysics,
{\bf 15}
  (1971) 337.

\bibitem{Vangioni-Flam90}
E.~Vangioni-Flam, M.~Cass\'{e}, J.~Audouze, Y.~Oberto: Astrophysical
Journal,
  {\bf 364} (1990) 568.

\bibitem{Webber98}
W.~R.~Webber et~al.: Phys. Rev., {\bf C 58} (1998) 3539.

\end{thebibliography}

\end{document}